\documentclass[aps,superscriptaddress,prb,twocolumn,floatfix]{revtex4}

\usepackage[utf8]{inputenc}
\usepackage[T1]{fontenc}
\usepackage[english, portuguese, english]{babel}

\usepackage{xcolor}
\definecolor{maincolor}{HTML}{862232}

\usepackage[]{url}
\usepackage[breaklinks=true]{hyperref}

\usepackage{indentfirst}
\usepackage[subpreambles=true]{standalone}

\usepackage{natbib}
\usepackage{braket}

\usepackage{graphicx}
\usepackage[]{wrapfig}
\usepackage{subfigure}

\usepackage{amsmath, amsfonts, amssymb, mathrsfs} 

\newcommand{\upS}{\uparrow} 
\newcommand{\downS}{\downarrow} 

\newcommand{\Tr}[1]{\text{Tr} \left[ #1 \right]}

\newcommand{\Nop}{\hat{n}}
\newcommand{\Cop}{\hat{c}}

\definecolor{blueMoon}{HTML}{6A0888}
\definecolor{sifscIIIa}{HTML}{7B03A9}
\definecolor{sifscIII}{HTML}{11bbaa}

\newcommand{\ba}{\begin{eqnarray}}
\newcommand{\ea}{\end{eqnarray}}

\newcommand{\be}{\begin{equation}}
\newcommand{\ee}{\end{equation}}

\usepackage{comment}

\begin{document}\sloppy

\title{
Work-distribution quantumness and irreversibility when crossing a quantum phase transition in finite time
}

\author{Krissia Zawadzki}
\affiliation{Department of Physics, Northeastern University, Boston, Massachusetts 02115, USA}

\author{Roberto M. Serra}
\affiliation{Centro de Ciências Naturais e Humanas, Universidade Federal do ABC,
Avenida dos Estados 5001, 09210-580, SantoAndré, São Paulo, Brazil
}

\author{Irene D'Amico}
\affiliation{Department of Physics, University of York, York, YO10\,5DD, United Kingdom}
\affiliation{Departamento de Física e Ciência Interdisciplinar, Instituto de Física de São Carlos,
  University of São Paulo, Caixa Postal 369, 13560-970 São Carlos, SP, Brazil}
\affiliation{International Institute of Physics, Federal University of Rio Grande do Norte, Natal, Brazil
}

\begin{abstract}
The thermodynamic behavior of out-of-equilibrium quantum systems in finite-time dynamics encompasses the description of energy fluctuations, which dictates a series of system’s physical properties. In addition, strong  interactions in many-body systems strikingly affect the energy-fluctuation statistics along a non-equilibrium dynamics. By driving transient currents to oppose the precursor to metal-Mott insulator transition in a diversity of dynamical regimes, we show how increasing correlations dramatically affect the statistics of energy fluctuations and consequently the quantum work distribution of finite Hubbard chains. Statistical properties of such distributions, as its skewness, that changes dramatically across the transition, can be related to irreversibility and entropy production. Even close to adiabaticity, the quasi quantum phase transition hinders equilibration, increasing the process irreversibility, and inducing strong quantum features in the quantum work distribution. In the Mott-insulating phase the work fluctuation-dissipation balance gets modified, with the irreversible entropy production dominating over work fluctuations. The effect of an interaction-driven quantum-phase-transition on thermodynamics quantities and irreversibility has to be considered in the design of protocols in small scale devices for application in quantum technology. Eventually, such many-body effects can also be employed in work extraction and refrigeration protocols at quantum scale.
\end{abstract}

\maketitle

\paragraph*{Introduction ---}
After more than a century, the well-established laws of thermodynamics have been challenged by the quantum nature of nanoscale systems\cite{Horodecki:2013, Kosloff:2013, Goold_2016, Vinjanampathy-CP.57.4.2016, binder2018thermodynamics,Millen:2016,Alicki:2018,Binder:2019};  quantum thermodynamics is now extending  concepts such as heat, work, and entropy\cite{Horodecki:2013, Skrzypczyk:2014, DelRio:2011} to this scale. At the same time, working conditions for quantum technology devices often correspond to finite-temperatures, non-equilibrium regimes, so that development of related formalism is in high demand. In quantum systems, thermodynamic probability distributions  contain rich information about the possible transitions between eigenstates \cite{Campisi:2017} and, more interestingly, thermal and quantum fluctuations \cite{Dorosz:2008, Batalhao:2014, Landi:2016}, equilibration and irreversibility \cite{Batalhao:2015, Hoang:2016,Zhong:2015, Apollaro:2015}. For small quantum systems these distributions are not symmetric and narrow Gaussian distributions \cite{Horodecki:2013}.

Quantum phase transitions (QPTs) are an exquisitely quantum
phenomenon, so there is interest in investigating their signature on quantum thermodynamic quantities and their distributions \cite{ Sindona:2014, Marino:2014, Apollaro:2015,  Mascarenhas:2014,Xu:2018, Wang:2018, Hickey:2014,Landi:2016,Hoang:2016,Schmidtke:2018, Li_2019}. In addition, many-body interactions, which are ubiquitous and notoriously difficult to treat, assume an even more complex role in out-of-equilibrium quantum systems  \cite{Dorner:2012, Arrais:2019}, where, e.g., they may affect the way the system reaches or settles into different phases. Relevant questions are:
  what is the role of many-body correlations for quantum particles driven out of equilibrium, and how do they affect quantum
thermodynamical quantities? Do they contribute or oppose reversibility \cite{Micadei:2017} and thermalization ? What if many-body correlations induce a QPT, what signatures appear in  thermodynamic distributions? And how do they depend on the system size?

Most of the previous studies of QPT signatures in quantum
thermodynamics focused on  QPTs driven by external fields and/or on the sudden quench regime.
They analysed features of  quantum thermodynamic quantities, sometimes up to the second moment of their distribution, and  their evolution as the critical parameter, usually an external field, is (suddenly) driven across the transition.

In this Letter, we consider the above questions in the context of microscopic models for strongly correlated systems undergoing {\it finite time} processes at finite temperature. We study the non-homogeneous  one-dimensional Hubbard model at half filling, as it is driven out of equilibrium. Finite Hubbard chains may undergo a precursor to the metal-Mott insulator transition, a QPT driven solely by many-body interactions.
 Our focus is the quantum work probability distribution and its statistics:  we inspect the first three moments, related to the mean, variance, and skewness. The latter has been to a large extent overlooked, and we demonstrate that it allows us to characterize the transition between the different coupling regimes, including the precursor to the metal-Mott insulator QPT (pM-QPT), as well as the different dynamical regimes (sudden quench to nearly-adiabatic).
Our results also demonstrate that by considering the
sudden quench regime alone, one misses the contribution of the dynamics to the QPT signatures, which becomes dominant in
finite-time regimes.
Many-body interactions strikingly affect the shape of the quantum work probability distribution: while it acquires some classical features for increasing system size and weak correlations, these are completely dismantled by the pM-QPT, which also averts the system from equilibrium. Interestingly, we show that, in the Mott-insulating phase, entropy production dominates over work fluctuations, in contrast to the literature\cite{Hermans:1991,Wood:1991,Miller:2019}.
Finally, we relate the skewness with the entropy production, and propose its role as a witness of irreversibility for many-body systems out-of-equilibrium.

\paragraph*{Driven Hubbard chains ---}
The
Hubbard model allows for both itinerant electron spins (conduction band) and
localized magnetic moments. It was initially designed to describe
strongly correlated systems such as transition metals; more recently it has
been utilized to describe systems of importance to quantum technologies, such
as cold atoms in an optical lattice, chains of trapped ions, excitons and electrons in coupled quantum dots, or small molecules \cite{Drewes:2016, Drewes:2017, Boll:2016, Braun:2014, Sherg:2018, Coe:2011}.
Even non-driven, short Hubbard chains are characterized by a very rich
physical behaviour, with many-body interactions driving  a precursor to the
 metal to Mott insulator transition, \cite{Greentree:2006, Murmann:2015, Zawadzki:2017} and studies of a driven Hubbard dimer show promising results\cite{Carrascal:2015,Herrera:2017}.

Here we consider half-filled fermionic chains undergoing
a process in which a time-dependent electric field is applied for a finite-time. Their  Hamiltonian is
	\ba
		\label{eq:Hubbard_driven}
		H(t) & = & -J \sum_{j=1}^{L-1} (\Cop_{j, \sigma}^\dagger \Cop_{j+1, \sigma} + \Cop_{j+1, \sigma}^\dagger \Cop_{j, \sigma})
		+ U \sum_{j=1}^{L} \Nop_{j\upS} \Nop_{j\downS} \nonumber\\ & + & \sum_{j=1}^{L} V_j(t) \Nop_{j\sigma},
	\ea
	where, $\Cop_{j\sigma}^\dagger$ ($\Cop_{j\sigma}$) are the creation (annihilation) operators for a fermion with spin $\sigma = \upS, \downS$ in the $j$-th site,
	$n_{j\sigma} = \langle \Cop_{j\sigma}^\dagger \Cop_{j\sigma} \rangle$ represents the corresponding $j$-site occupation, $J$ is the hopping parameter, $U$ is the Coulomb on-site repulsion, and $V_j(t) =  \Delta_j \, t / \tau $, with $\Delta_j=\frac{10J}{L-1} j$, is the time-dependent linear  potential that drives an out-of-equilibrium transient current along the chain.

The system is initially in thermal equilibrium at temperature $\beta^{-1}=k_BT = 2.5J$ (where $k_B$ is the Boltzmann constant and $T$ the absolute temperature), with $\rho(t=0)  = {e^{-\beta H(t=0)}}/{Z_{t=0}}$, and $Z_{t=0} = \Tr{e^{-\beta H(t=0)}}$. The driving time $\tau$ controls the rate of the dynamics that steers $H_0=H(t=0)$ to $H_f=H(t=\tau)$. The final Hamiltonian $H_f$ is independent of $\tau$.
Our results were obtained via exact diagonalization;
the time-evolution calculated by a routine provided by
the QuTip package \cite{QuTip}.

\paragraph*{Statistics of work and correlations ---}
The probability distribution characterizing the quantum work \cite{Skrzypczyk:2014} performed on the closed system\footnote{We consider processes fast enough to be represented as (a unitary) closed system dynamic. In other words, the calculations correspond to the scenario where the time duration of the driving protocol, $\tau$, is much smaller than any decoherence or relaxation times.} is given by
\begin{align}
		\label{eq:PW}
		P(W) & = \sum_{n, m} p_{n}^{0} p_{m|n}^{\tau} \delta[W - (\epsilon_m^\tau - \epsilon_n^0)],
	\end{align}
where  $p_{n}^{0}$ is the initial-state occupation probability of the n-th eigenstate $\ket{n}$ of energy $\epsilon_n^0$
of $H_0$,
and
$p_{m|n}^{\tau}$ is the conditional probability for $\ket{n}$ to make a transition to the
m-th eigenstate $\ket{m}$ of $H_f$.

\begin{figure}	
		\includegraphics[width=1\columnwidth]{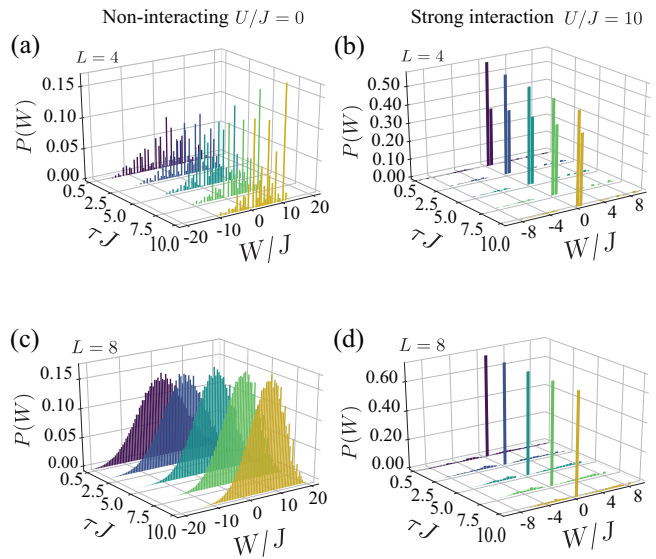}
		\caption{Quantum work distribution $P(W)$ for fermionic Hubbard chains at half filling driven by a time-dependent electric potential. Panels (a,b) refer to $4$-site chains, whereas (c,d) to 8 sites. The left panels (a,c) show the non-interacting case ($U = 0$) and the right panels (b,d) the strong-interaction regime ($U = 10 J$). Each panel displays $P(W)$ for different  driving times, from sudden-quench ($\tau=0.5/J$) to a close-to-adiabatic ($\tau=10/J$) dynamics.
}%
		\label{fig1}%
	\end{figure}

The complexity of $P(W)$  scales with the number of the possible energy transitions. In the systems we consider, half-filling with zero magnetization, the number of allowed transitions increases from 16 for $L=2$, to $5\times 10^3$ for $L=8$. \footnote{There are 4900 spin configurations for $L=8$ having $S_z = 0$.}  This is highlighted by Figs.~\ref{fig1}a and \ref{fig1}c, where $P(W)$ is shown for $L=4$ and $L=8$ for  the non-interacting case ($U=0$). The exponential increase in the number of transitions transforms the distribution from an irregular set of peaks to a bell shape;
changes in the type of dynamics -- from sudden quench ($\tau =0$) to close-to-adiabatic behaviour ($\tau = 10/J$) -- strongly affect the shape of the distribution, which  becomes increasingly asymmetric as $\tau$ increases.  On the contrary, when considering the strongly interacting regime ($U=10J$, Figs.~\ref{fig1}b and \ref{fig1}d) the shape
of $P(W)$ seems basically unaffected\cite{supplemental}. We attribute this behaviour to the insulating phase which de-facto substantially reduces the available Hilbert space.

This qualitative picture is quantified by the $k$-th  central moments of the quantum work distribution $P(W)$,
	\begin{align}
		\label{eq:kth_central_moment}
		\bar{\mathcal{W}}^k & = \langle (W-\bar{W})^k \rangle = \sum_{i} \, P(W_i) (W_i-\bar{W})^k.
	\end{align}

	\begin{figure}[htb!]

\centering
        \includegraphics[width=0.925\columnwidth]{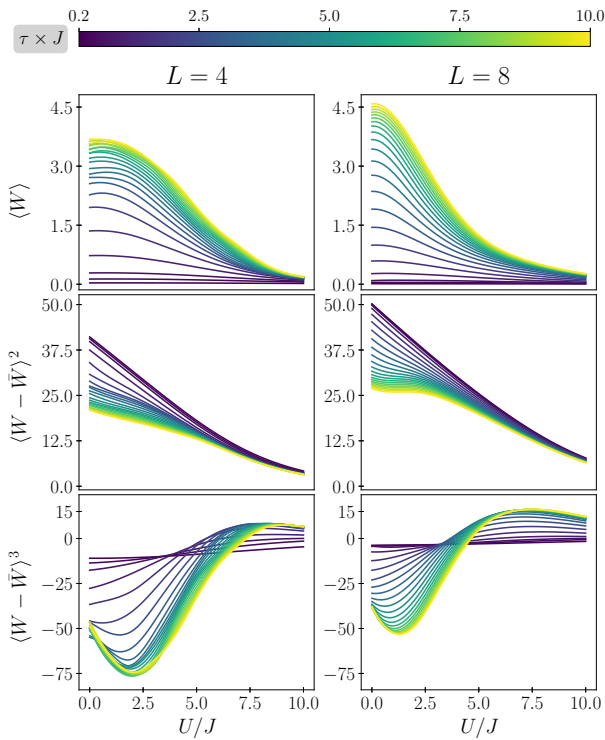}
		\caption{First three moments of the quantum work distribution (as labelled) versus $U$, for $0.2/J\le\tau\le 10 /J$,
and chain length $L=4$  (left) and $L = 8$ (right).
}
    \label{fig2}
	\end{figure}	

\begin{figure}[htb!]
	
		\centering
        \includegraphics[width=0.95\columnwidth]{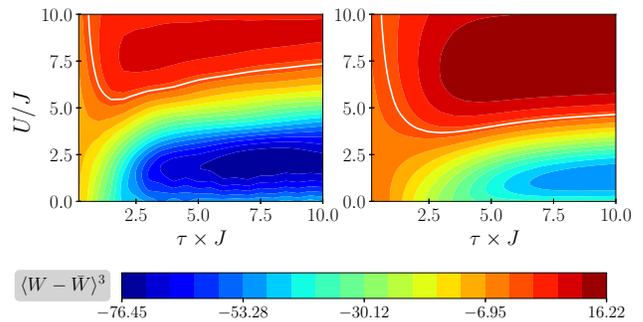}

		\caption[]{Heatmaps of the skewness of the quantum work distribution,  for $L = 4$ (left) and $L=8$ (right). The white line indicates $\bar{\mathcal{W}}^3=0$.
        }%
		\label{fig3}%
	\end{figure}

The moments $k=1$ (mean), $k=2$ (variance), $k=3$ (skewness) are shown in Fig.~\ref{fig2}, $L=4$ left and $L=8$ right; the corresponding `heatmaps' for  $k=3$ in  Fig.~\ref{fig3}, where the white line indicates $\bar{\mathcal{W}}^3=0$ \cite{supplemental}. The first three moments are
strongly dependent on $\tau$ for weak interactions, $U\approx 0 $, while almost $\tau$-independent for $U\approx 10 J$, once interactions have driven the pM-QPT
and the system becomes insulating. Regardless of the huge increase in the Hilbert space, the behaviour across the transition is qualitatively independent from the system size, hinting to a possible scaling behaviour.
The most
striking features appear in the skewness $\bar{\mathcal{W}}^3$.
For sudden quenches, $\tau \ll J^{-1}$, the skewness is relatively small and depends only weakly on $U$ (see Fig.~\ref{fig3}).  However, for finite-time processes,
 $\tau\stackrel{>}{\sim}0.5/J$, $\bar{\mathcal{W}}^3$ changes sign across the pM-QPT (white line in Fig.~\ref{fig3}), with proper minima and maxima bracketing the transition when $\tau\stackrel{>}{\sim}2.5/J$ (see Fig.~\ref{fig3} and Fig.~\ref{fig2}, lower panels).
As $U$ increases,
 the system suffers a dynamic competition between the  transient current induced by the drive and the increasing on-site repulsion. This leads to  a dramatic change in the shape of $P(W)$, with a marked asymmetry shifting from left (before pM-QPT) to right (after pM-QPT). As $\tau$ increases, the region in-between $\bar{\mathcal{W}}^3(U)$ extrema shifts towards larger $U$'s (see Fig.~\ref{fig2}, lower panels).
 In classical
thermodynamics, probability distributions tends to be symmetric, and so it is particularly fit that a strong asymmetry in the distribution, and a dramatic change of this asymmetry,  signals an exquisitely quantum phenomenon such as a QPT.

\paragraph*{Entropy production and irreversibility ---} Together with the statistics of work, we can inspect how the pM-QPT affects irreversibility. We quantify this by considering the entropy production
\ba
		\label{eq:entropy production}
		\left\langle \Sigma \right\rangle 
		& = & S \left( \rho_\tau  ||\rho_\tau^{\text{eq}} \right),
	\ea
where, $S \left( \rho_\tau  ||\rho_\tau^{\text{eq}} \right)=\text{Tr} \rho_\tau \left(\ln\rho_\tau - \ln\rho_\tau^{\text{eq}} \right)$
defines the Kullback relative entropy between the final state $\rho_\tau=\mathcal{U}_{\tau} \rho_0^{\text{eq}} \mathcal{U}_{\tau}^\dagger$,  and its equilibrium counterpart
 $\rho_\tau^{\text{eq}} = {e^{-\beta H(t=\tau)}}/{Z_{t=\tau}}$, with $\mathcal{U}_t$ the time-evolution operator.
 We note that $\left\langle \Sigma \right\rangle/ \beta$ corresponds also to the energy that would be dissipated if thermalization would follow the finite-time driven protocol. We examine the entropy production in our systems in various dynamical and coupling regimes, full results for $L=4$ and $L=8$ are reported in the Supplemental Material. \cite{supplemental}

\begin{figure}[htb!]		
	\centering

		\includegraphics[width=1\columnwidth]{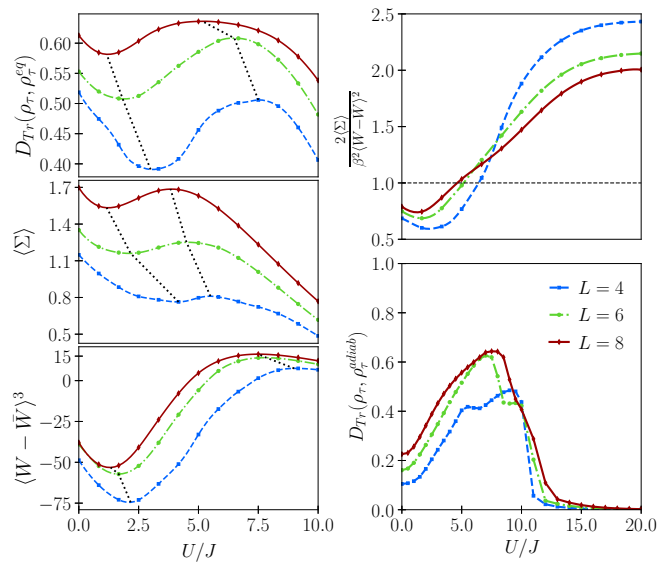}
		\caption{Left panels: Trace distance $D_{\text{Tr}}(\rho_\tau, \rho_\tau^{\mathrm{eq}})$(top),
		entropy production $\langle \Sigma \rangle $ (middle), and skewness $\langle W - \bar{W} \rangle^3$ (bottom), versus coupling strength $U/J$ and
		for chains of size $L=4,6,8$ and $\tau \times J = 10$. The dashed black lines connect minima and maxima for increasing system size.\\
	 Right panels:  entropy production to work fluctuations ratio  versus coupling strength $U/J$,
		for chains of size $L=4,6,8$ and $\tau \times J = 10$ (top).  Trace distance $D_{\text{Tr}}(\rho_\tau, \rho_\tau^{\mathrm{adiab}})$ between final and corresponding adiabatic state, same parameters as upper panel (bottom).
}
    \label{fig5}
	\end{figure}

For a finite quantum system, adiabaticity does not imply equilibration, hence, to quantitatively investigate this discrepancy,  we focus on large $\tau$ results, and use, in addition to  $\left\langle \Sigma \right\rangle$, the trace distance between final and corresponding equilibrium state, $D_{Tr}(\rho_\tau,\rho^{\mathrm{eq}}_\tau)= Tr \left[ \sqrt{\left(\rho_\tau -\rho^{\mathrm{eq}}_\tau\right)^\dagger \left(\rho_\tau -\rho^{\mathrm{eq}}_\tau\right)
}\right]/2$.  This is plotted in the top left panel of Fig. \ref{fig5} together with the entropy (middle left) and the skewness (bottom left) as a function of $U/J$, for $\tau \times J = 10$ and $L=4,6,8$. We find that these quantities seem to display a scaling trend with system size. In particular, all quantities similarly signal the pM-QPT, moving from a minimum to a maximum. These extrema all shift towards $U = 0^+$ (the thermodynamic limit for the metal-Mott insulator QPT) as $L$ increases (see dashed lines in Fig.~\ref{fig5}, left panels).
The transition pulls the final state away from equilibrium as demonstrated by the corresponding increase of $D_{Tr}(\rho_\tau,\rho^{\mathrm{eq}}_\tau)$, and dramatically affects the work distribution shape, as witnessed by the change in sign of the skewness. After the pM-QPT, as interactions increase further, the final state draws nearer to equilibrium, as the system, now almost an insulator, poorly responds to the applied field. Indeed, the work distribution comprises here very few transitions (Fig.~\ref{fig1}b and d).

We note that the value of the trace distance demonstrates that the final system remains always significantly far from equilibrium,
even when the skewness is zero ($U/J\approx 5$) and the distribution becomes more akin to the linear response form,
$\langle W \rangle =  \Delta F + \frac{\bar{\mathcal{W}}^2}{2k_B T}$,
which is valid close to equilibrium, with  $\Delta F = \langle W \rangle -\langle \Sigma\rangle/\beta$  the free energy variation.

Fig.~\ref{fig5} shows that increasing the system size drives a system farthest from equilibrium and increases the quantity of energy  $\langle\Sigma\rangle/\beta$ to be dissipated for reaching it. However, the system size affects the work distribution asymmetry in opposite ways before and after the quasi-QPT. Before the pM-QPT, availability of an exponentially increasing number of transitions `regularize' the distribution (compare Fig.~\ref{fig1}a and c) contributing to the decrease of its asymmetry, while, by de-facto restricting the available Hilbert space, the pM-QPT restores full quantum features in $P(W)$, even for increasing size (compare Figs.~\ref{fig1}b and d).

\paragraph*{Entropy production and work fluctuation-dissipation relation}
Close to adiabaticity, classical processes satisfy the work fluctuation-dissipation relation
$\left\langle \Sigma \right\rangle = \beta^2\bar{\mathcal{W}}^2/2$\cite{Hermans:1991,Wood:1991}; however, recent studies\cite{Miller:2019}
suggest that, for slow quantum processes in open systems, this is governed by the inequality
	\begin{align}
		\label{eq:fl-diss}
\left\langle \Sigma \right\rangle \le \beta^2\bar{\mathcal{W}}^2/2.
\end{align}
We examine the effect of the pM-QPT on the work fluctuation-dissipation relation in Fig.~\ref{fig5}, right upper panel, and show that the transition is marked by a {\it reversing} of the inequality (\ref{eq:fl-diss}), with work fluctuations hence becoming smaller than dissipation.  Most interestingly, after the pM-QPT, while  increasing $U$ leads the dynamical process back to adiabaticity (Fig.~\ref{fig5}, right lower panel, $U>10J$), dissipation remains dominant over work fluctuations, even for very small values of $D_{\text{Tr}}(\rho_\tau, \rho_\tau^{\mathrm{adiab}})$. This reversing of (\ref{eq:fl-diss}) is a many-body effect: the pM-QPT dramatically reduces the system response to the applied field, and hence the width of the work distribution, for all rate of driving, including slow driving (see Fig.~\ref{fig2}, middle panels).

\paragraph*{Conclusion ---}
We discussed the effects of many-body interactions on the statistics of work in fermionic chains driven for finite times. We considered dynamics from sudden quench to quasi-adiabaticity, and observed the signatures of the precursor to the metal-Mott insulator  quantum phase transition.
Our results show that, when the system is weakly correlated, the work probability distribution $P(W)$ is highly sensitive to the rate of driving, whereas it remains almost unaffected when many-body interactions are strong.

 If the chains' length $L$ is increased and $U/J \stackrel{<}{\sim}1$,  $P(W)$  acquires some of the features of a classical distribution, such as
a well-defined maximum and a bell shape.
In contrast, after crossing the precursor to the QPT, for $U/J \stackrel{>}{\sim}5$, the quantum nature of the system dominates at all the explored values of $L$, strongly hindering work extraction with, nonetheless, a price paid in a residual entropy production.
The quasi-Mott-insulating phase is associated with a striking reduction of the number of transitions arising from the dynamics, so that $P(W)$ becomes almost independent on the rate of variation of the external field. This feature leads to entropy production dominating work fluctuations even for slow processes, in contrast to the classical work fluctuation-dissipation relation, and at difference with recent predictions for slowly-driven open quantum systems.

For dynamics beyond sudden quenches, a change in sign and a remarkable variation in value of the skewness characterize the precursor to the metal-Mott insulator transition. These features persist even when
the number of degrees of freedom is exponentially increased.
In the sudden quench regime, the precursor to the QPT affects $P(W)$ only through its effects
on the initial and final Hamiltonians' eigenstates; instead, for finite driving times, the precursor to the metal-to-Mott insulator
transition affects $P(W)$ twice, through its effect on the eigenstates and by modifying the system response to the applied drive. This leads to {\it qualitatively} different signatures of the precursor to the QPT on the quantum work distribution, depending on the dynamical regime.

By comparing to the trace distance between the final and the corresponding equilibrium state, we conclude that the third moment of $P(W)$ also retains information about the entropy production and equilibration across the precursor to the QPT.

Experimental realizations of correlated quantum matter could be implemented by means of small molecules and NMR \cite{Batalhao:2015,Micadei:2017}, coupled quantum dots and ion traps
\cite{Rosnagel:2016,Shuoming:2015},
or cold atoms platforms \cite{Chiara:2015,Cerisola:2017}. Our findings may help to design time-dependent protocols which exploit many-body interactions for, e.g., tailoring work extraction or optimizing efficiency of a refrigeration cycle where the coolant is a strong correlated many-body system, yielding to novel applications of quantum thermodynamics

We  thank Marcela Herrera for very usefull discussions during the early stages of the work.
We acknowlege financial support from the Royal Society (grant no. NA140436); CNPq (grant no. PVE-401414/2014-0); CAPES (grant no. PDSE-88881.135185/2016-01); (INCT-IQ). KZ acknowledges the Schumblenger Foundation for sponsorship through the program Faculty for the Future.

\Urlmuskip=0mu plus 1mu\relax
\bibliography{references_QT_no_http}

\end{document}